\documentclass[3p,times,twocolumn]{elsarticle}
\newcommand{\lsi}{\,\raisebox{-0.13cm}{$\stackrel{\textstyle<}
{\textstyle\sim}$}\,}

\usepackage{ecrc}

\volume{00}

\firstpage{1}

\journalname{Nuclear Physics B Proceedings Supplement}

\runauth{}

\jid{nuphbp}

\jnltitlelogo{Nuclear Physics B Proceedings Supplement}

\usepackage{amssymb}

\usepackage[figuresright]{rotating}

\begin{document}

\begin{frontmatter}



\dochead{}

\title{Fermi-LAT measurement of the diffuse gamma-ray emission and constraints on the Galactic Dark Matter signal}


\author{G.~Zaharijas}
\address{International Center for Theoretical Physics, Trieste, Italy}
\address{Institut de Physique Th\'eorique, CEA/Saclay, F-91191 Gif sur Yvette, France}
\author{J.~Conrad, A.~Cuoco, Z.~Yang}
\address{Department of Physics, Stockholm University, AlbaNova, SE-106 91 Stockholm, Sweden}
\address{The Oskar Klein Centre for Cosmoparticle Physics, AlbaNova, SE-106 91 Stockholm, Sweden}

\author{(for the {\it Fermi} LAT collaboration)}

\begin{abstract}
We study diffuse gamma-ray emission at intermediate Galactic latitudes measured by the Fermi Large Area Telescope with the aim of searching for a signal from  dark matter annihilation or decay. In the absence of a robust dark matter signal, 
we set conservative dark matter limits requiring that the dark matter signal does not exceed  the observed diffuse gamma-ray emission. A second set of more stringent limits is derived based on modeling the foreground astrophysical diffuse emission. 
Uncertainties in the height of the diffusive cosmic-ray halo, the distribution of the cosmic-ray sources in the Galaxy, the cosmic-ray electron index of the injection spectrum and the column density of the interstellar gas are taken into account using a profile likelihood formalism, while the parameters governing the cosmic-ray propagation have been derived from fits to local cosmic-ray data. 
The resulting limits impact the range of particle masses over which dark matter thermal production in the early Universe is possible, and challenge the interpretation of the PAMELA/Fermi-LAT cosmic ray anomalies as annihilation of dark matter.
\end{abstract}


\end{frontmatter}


\section{Introduction}
The Milky Way halo has long been considered a good target for searches of indirect signatures of dark matter (DM). WIMP DM candidates are expected to produce gamma rays, electrons and protons in their annihilation and decays and such emission originating in our Galaxy would appear as a diffuse signal. At the same time, the majority of the Galactic diffuse emission is produced through radiative losses of cosmic-ray (CR) electrons and nucleons in the interstellar medium. Modeling of this emission presents one of the major challenges when looking for subdominant signals from dark matter.

In this analysis we test the diffuse LAT data for a contribution from the DM signal by performing a fit of the spectral and spatial distributions of the expected photons at intermediate Galactic latitudes. In doing so, we take into account the most up-to-date modeling of the established astrophysical signal, adapting it to the problem in question \cite{us}.
Our aim is to constrain the DM properties and treat the parameters of the astrophysical diffuse gamma-ray background as nuisance parameters. 
Besides  this approach,  we will also quote conservative upper limits using  the data only (i.e.  without performing any modeling of the astrophysical background). 

We follow \cite{paper2} in using the \texttt{GALPROP} code v54 \cite{galprop}, to calculate the propagation and distribution of CRs in the Galaxy and the whole sky diffuse emission. 
In \cite{paper2} various standard parameters of the CR propagation were studied in a fit to CR data 
and it was shown that they represent well the gamma-ray sky, although various residuals ({at a $\sim 30\%$ level \cite{paper2}}), both at small and large scales, remain. 
In our work, we use the results of the fits to the CR data from \cite{paper2} but we allow for more freedom in certain parameters governing the CR distribution and known astrophysical diffuse emission and constrain these parameters by fitting the models to {\it the LAT gamma-ray data}. Despite the large freedom we leave in the models 
we see residuals in our ROI at the $\pm 30\%$ level and at $\sim 3~\sigma$ significance. These residuals can be ascribed to various limitations of the models: imperfections in the modeling of gas and ISRF components, simplified assumptions 
in the propagation set-up, 
unresolved point sources, 
and large scale structures like Loop I \cite{Casandjian:2009wq} or the Galactic Bubbles \cite{Su:2010qj}. Since residuals do not seem obviously related to DM, we focus in the following on setting limits on the possible DM signal, rather than \emph{searching} for a DM signal.

\section{DM maps}

 We parametrize the  smooth DM density $\rho$ with a NFW spatial profile \cite{Navarro:1995iw} 
 \begin{equation}
 \rho(r)=\frac{\rho_0\,R_s}{r \, \left(1+r/R_s  \right)^{2}}
  \end{equation}
  and a cored (isothermal-sphere) profile \cite{Begeman:1991iy}: 
  \begin{equation}
  \rho(r) = \frac{\rho_0 \left( {R_\odot^2+R_c^2}\right)}{\left({r^2+R_c^2}\right)}.
   \end{equation}
  For the local density of DM we take the value of $\rho_0=0.43$ GeV cm$^{-3}$ \cite{Salucci:2010qr}, and the scale radius of $R_s=$ 20 kpc (for NFW) and $R_c=$ 2.8 kpc (isothermal profile). We also set the distance of the solar system from the center of the Galaxy  to the value $R_\odot=$ 8.5 kpc.
For the annihilation/decay spectra we consider three channels with distinctly different signatures: annihilation/decay into the $b{\bar b}$ channel,  into $\mu ^+ \mu^-$, and into $\tau^+\tau^-$. In the first case gamma rays are produced through hadronization of annihilation products and subsequent pion decay. The resulting spectra are similar for all channels in which DM 
produces heavy quarks and gauge bosons and this channel is therefore representative for a large set of  particle physics models. 
The choice of leptonic channels provided by the second and third scenarios, 
is motivated by the dark matter interpretation  \cite{Grasso:2009ma} of the PAMELA  positron fraction \cite{Adriani:2008zr} and the {\it Fermi} LAT  electrons plus positrons \cite{Abdo:2009zk}  measurements. 
In this case, gamma rays are dominantly produced through radiative processes of electrons, as well as through the Final State Radiation (FSR). We produce the DM maps with a version of \texttt{GALPROP}  slightly modified to implement custom DM profiles and injection spectra (which are calculated by using  {the \texttt{PPPC4DMID} tool} described in \cite{Cirelli:2010xx}).

\section{Approach to set  DM limits} \label{outline}

We use 24 months of LAT data in the energy range between 1 and 100 GeV (but, we use energies up to 400 GeV when deriving DM limits with no assumption on the astrophysical background). We use only events classified as gamma rays in the P7CLEAN event selection and the corresponding \verb"P7CLEAN_V6" instrument response functions (IRFs)\footnote{http://fermi.gsfc.nasa.gov/ssc/}.  Structures like Loop I and the Galactic Bubbles appear mainly at high Galactic latitudes and to limit their effects on the fitting 
we will consider a ROI in Galactic latitude, $b$, of $5^{\circ} \leq |b|\leq 15^{\circ}$, and Galactic longitude, $l$, $|l|\leq 80^{\circ}$. We mask the region $|b|~\lsi 5^{\circ}$ along the Galactic Plane, in order to reduce the uncertainty due to the modeling of the astrophysical and DM emission profiles.

\subsection{DM limits with no assumption on the astrophysical background}\label{nobkg}
To set  these type of limits we first convolve a given DM model with the Fermi IRFs 
to obtain the counts expected from DM annihilation. 
The expected counts are then compared with the observed counts in our ROI and the upper limit is set to the \emph{minimum} DM normalization which gives counts in excess of the observed ones in at least one bin, i.e. we set $3\sigma$ upper limits given by the requirement  $n_{i DM}-3\sqrt{n_{i DM}} > n_i$, where $n_{i DM}$ is the expected number of counts from  DM in the bin $i$ and $n_i$ the actual observed number of counts.

\subsection{DM limits with modeling of astrophysical background}\label{main}
In this analysis we model the diffuse emission as a combination of a dark matter and a parameterized conventional astrophysical signal and we derive the limits on the DM contribution using {\it the profile likelihood method}. More precisely, for each DM channel and mass the model which describes the LAT data best is the one which maximizes the likelihood function defined as a product running over all spatial and spectral bins $i$ ,
\begin{equation}
 L_{k}(\theta_{DM})= L_k(\theta_{DM}, \hat{\hat{\vec{\alpha}}})= max_{\vec{\alpha}} \prod_{i} P_{ik}(n_{i};\vec{\alpha},\theta_{DM}),
\label{leq}
\end{equation}
where $P_{ik}$ is the Poisson distribution for observing $n_i$ events in bin $i$ given an expectation value that depends on the parameter set ($\theta_{DM}$, $\vec{\alpha}$). $\theta_{DM}$ is the intensity of the DM component, 
$\vec{\alpha}$ represents the set of parameters which enter the astrophysical diffuse emission model as linear pre-factors to the individual model components (cf. equation \ref{eq:F} below), while $k$ denotes the set of parameters which enter in a non-linear way. 
We sample non-linear parameters of the astrophysical background on a grid (for computational efficiency). 


The linear part of the fit is performed with the \texttt{GaRDiAn} code in the following way. The CR source distribution (CRSDs) is a critical parameter for DM searches and we define a parametric CRSD as sum of {\it step functions} in Galactocentric radius $R$, and treat the normalization of each step as a free parameter {\it in the fit to gamma rays}\footnote{CRSDs are traditionally modeled from the direct observation of tracers of SNR and can be observationally biased.}. In order to have conservative and robust limits  
we set to zero the ${e,p}$ CRSDs in the inner Galaxy region, within 3 kpc of the Galactic Center. In this way, potential $e$ and $p$ CR sources which would be required in the inner Galaxy  will be  potentially compensated by DM, producing conservative constraints. 
The linear part of the fit is then performed by combining the sky maps produced in each ring of CRSD of the main components of the diffuse emission: the emission from $\pi ^0$ decay, bremsstrahlung and inverse Compton. 
Additionally an isotropic component arising from the extragalactic gamma-ray background and misclassified charged particles needs to be included to fit the {\it Fermi} LAT data. 
In summary, the various \texttt{GALPROP}  outputs are combined as:
\begin{eqnarray}\nonumber
     F & = &  \sum_{i}    \bigg\{ c_i^p ~\big( H^i_{\pi^0} +  {\sum_{j}}~X_{\rm CO}^j H_{2 \ \! {\pi^0} }^{ij} \big) +         \\ \nonumber         
       & &    c_i^e ~ \big( H^i_{\rm bremss} +  {\sum_{j}}~X_{\rm CO}^j H_{2\ \! {\rm bremss}}^{ij}  + IC^i \big) \bigg\}  +          \\ \label{fiteq}
       & &   \alpha_\chi~(\chi_\gamma + \chi_{ic}) +  \sum_{m} \alpha_{IGB,m}~IGB^m .
       \label{eq:F}
\end{eqnarray}
The sum over $i$ is the sum over all step-like CRSD functions, the sum over $j$ corresponds to the sum over all Galactocentric annuli (details of the procedure of a placement of the gas in Galactocentric annuli and their boundaries are given in \cite{paper2}). $H$ denotes the gamma-ray emission from atomic and ionized interstellar gas
while $H_2$ the one from molecular hydrogen\footnote{It should be noted that in our case, where we mask $\pm 5^{\circ} $  along the plane, the expression actually simplifies considerably 
 since only the local ring $X_{\rm CO}$ factor enters the sum, since all the other $H_2$ rings do not extend further than 5 degrees from the plane.} and $IC$ the Inverse Compton emission.
$\chi_\gamma$ and $ \chi_{ic}$ are the prompt and Inverse Compton (when present) DM contribution and $\alpha_\chi$ the overall DM normalization. $\alpha_{IGB,m}$ and $IGB^m$ denote the Isotropic Gamma-ray Background (IGB) intensity for each of  the five energy bins over which the index $m$ runs.  In all the rest of the expression the energy index $m$ is implicit since we don't allow for the freedom of varying the \texttt{GALPROP}  output from energy bin to energy bin. We do not include sources in the fit as we use a mask to filter point sources from the 1FGL catalog \cite{1FGL}.

The outlined procedure is then repeated for each set of values of the non-linear propagation and injection parameters to obtain the full set of  profile likelihood curves. We scan over the three parameters: electron injection index, the height of the diffusive halo and the gas to dust ratio which parametrizes different gas column densities (see Table \ref{tab:summarypar}).
In this way we end up with a set of $k$  profiles of likelihood $L_k(\theta_{DM})$, one for each combination of the non-linear parameters. The envelope of these curves then approximates the \emph{final} profile likelihood curve, $L(\theta_{DM})$, where all the parameters, linear and non-linear have been included in the profile.
Limits are calculated from the profile likelihood function by finding the  $\theta_{DM,lim}$ values for which $L(\theta_{DM,lim})/L(\theta_{DM,max})$ is $\exp(-9/2)$ and $\exp(-25/2)$, for $3$ and 5 $\sigma$ C.L. limits, respectively. 

\section{Results}\label{results}

\begin{figure*}[t] 
\begin{center}$
\begin{array}{cc}
\includegraphics[width=0.45\columnwidth]{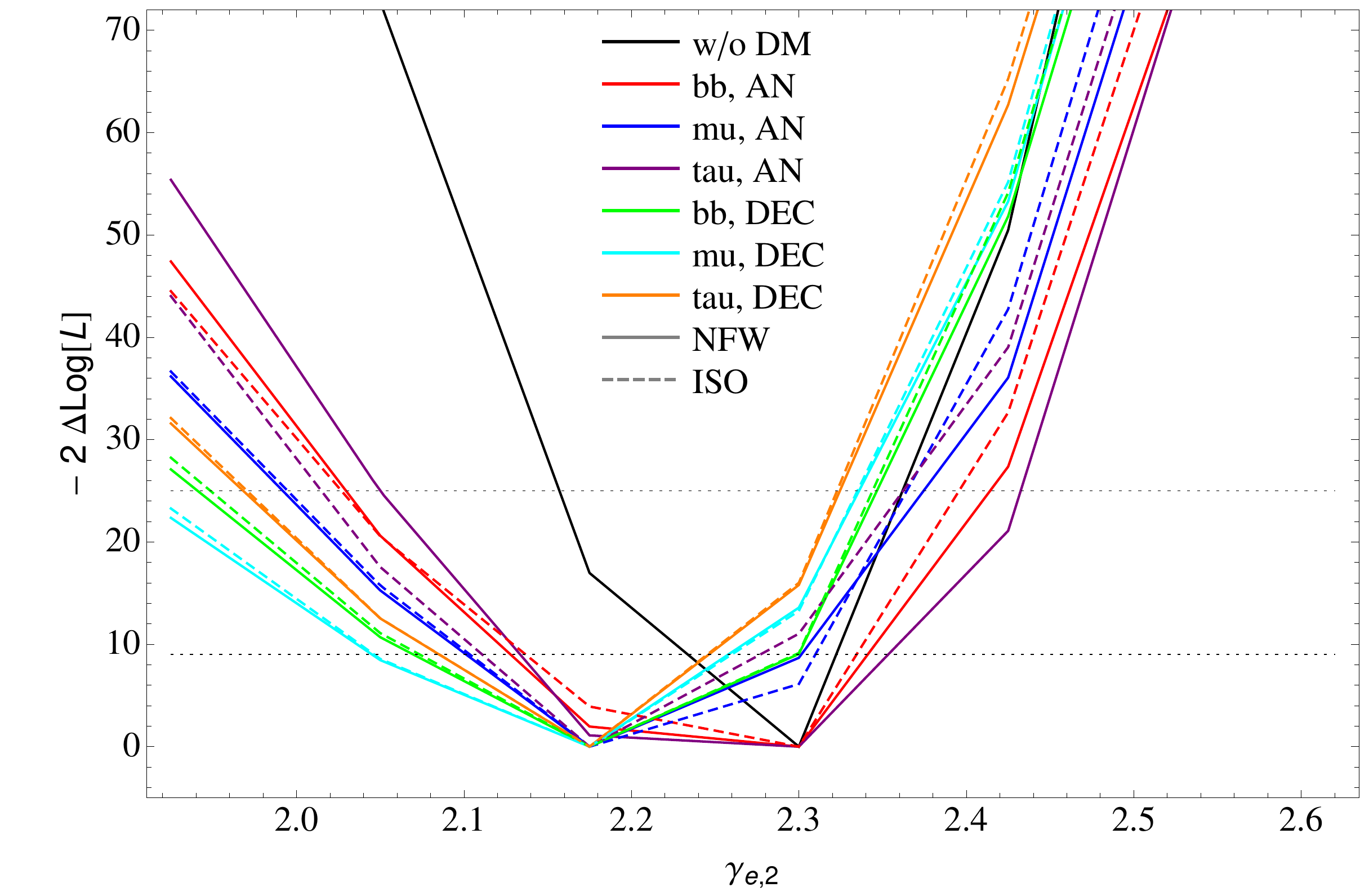}&
\includegraphics[width=0.45\columnwidth]{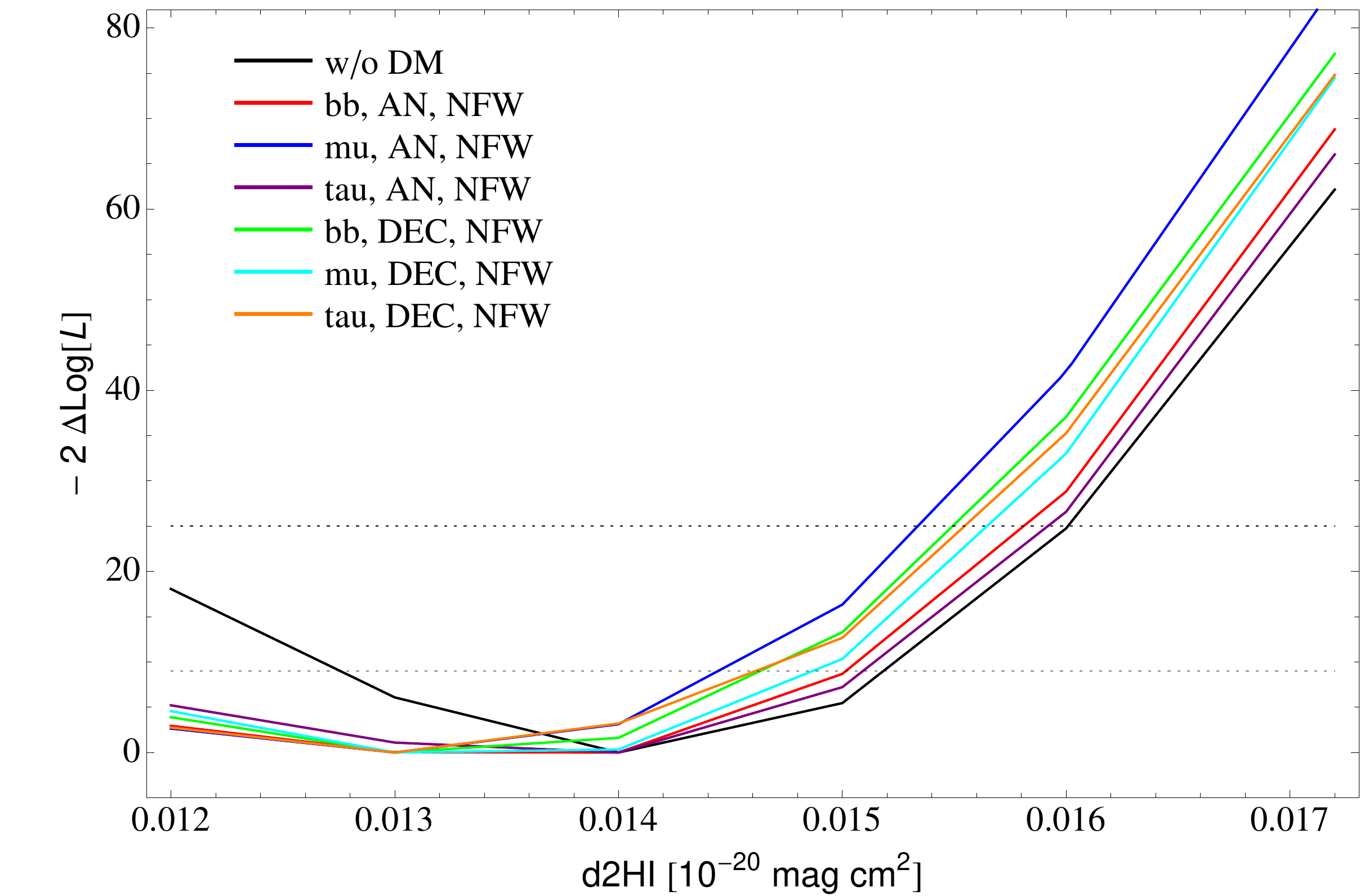}\\
\end{array}$
\end{center}
\begin{center}$
\begin{array}{c}
\includegraphics[width=0.45\columnwidth]{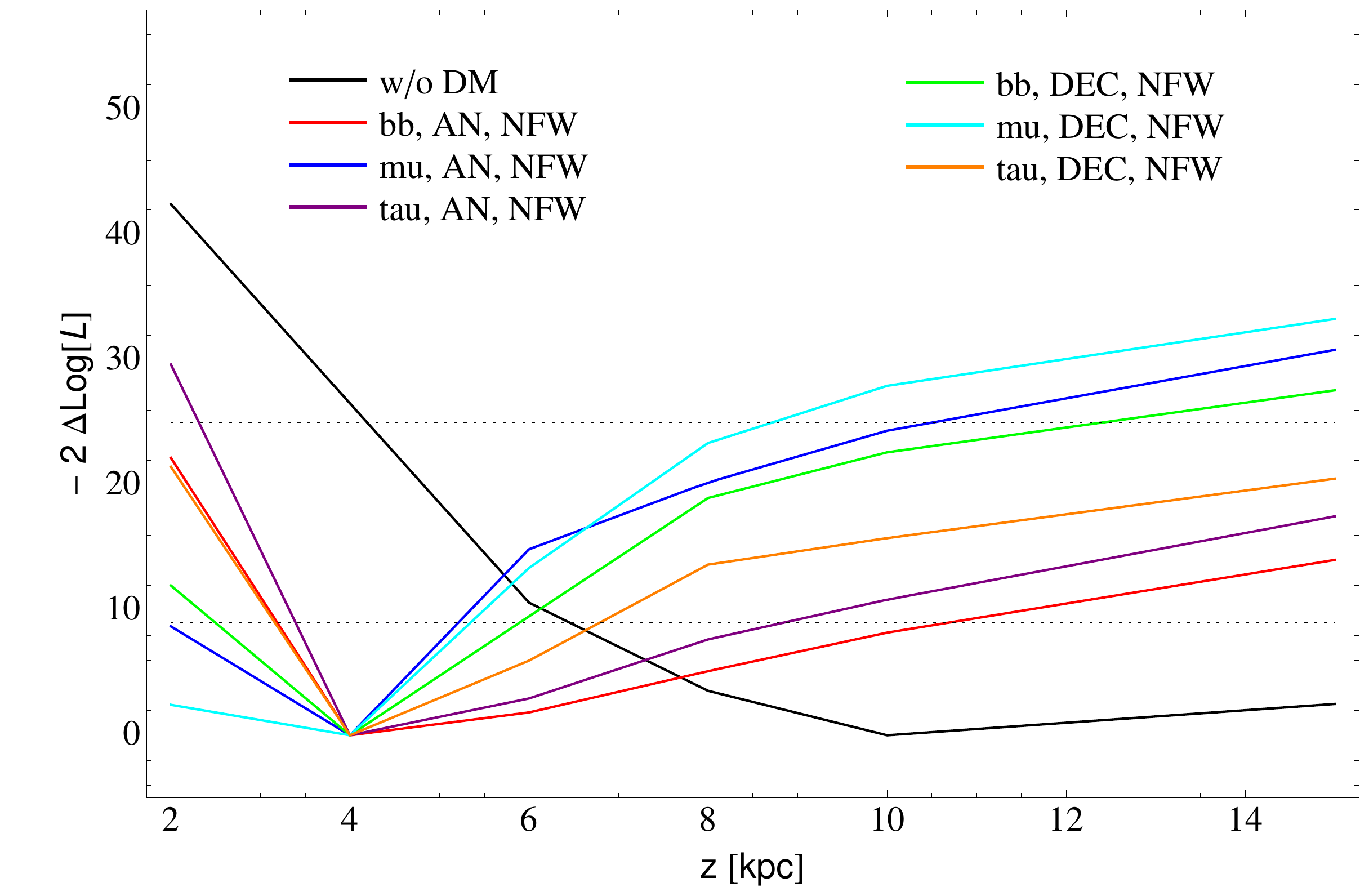}
\end{array}$
\end{center}
\caption{Profile likelihood curves for $z_h$, $\gamma_{e,2}$ and d2HI. The various curves refer to the case of no DM or different DM models (see the legend in the figure, where we mark a dominant decay (DEC) or annihilation (AN) channel and the assumed DM profile). All minima are normalized to the same level. Horizontal dotted lines indicate a difference in  $-2\Delta{\rm log}L$ from the minimum of 9 (3$\sigma$) and 25 (5$\sigma$).  \label{fig:profiles}}
\end{figure*}

An important point to note 
is that, for each DM model, the global minimum we found lies within the 3(5) $\sigma$ regions of many different models, providing a check against a bias in our procedure
. This point is illustrated in Figure~\ref{fig:profiles}, where the  profile likelihoods for the three nonlinear parameters, $z_h$, $\gamma_{e,2}$ and d2HI, are shown. 
To ease reading of the figure the profiling is actually performed with further grouping DM models with different DM masses, but keeping the different DM channels, DM profiles and the annihilation/decay cases separately.  The curve for the fit without DM is also shown for comparison. Each resulting curve has been further rescaled to a common minimum, since we are interested in showing that several models are within $-2\Delta{\rm log}L\leq 25$ around the minimum for each DM fit.
The $\gamma_{e,2}$ profile, for example, indicates that all models with $\gamma_{e,2}$ from 1.9 to 2.4 are within $-2\Delta{\rm log}L\leq 25$ around the minimum illustrating that the sampling around each of the minima for the six DM models is dense. Similarly, the d2HI profile indicates that all models with d2HI in the range  (0.120 - 0.160) $\times 10^{-20}$ mag cm$^2$ are within $5 \sigma$ from the minima for each of the six DM models. 
Finally the $z_h$ profile indicates that basically all the considered values of $z_h$ are close to the absolute minima. This last result is not surprising since, within our low-latitude ROI, we have little sensitivity to different $z_h$ and basically all of them fit equally well.
There is some tendency to favor higher values of $z_h$ when DM is not included in the fit, while with DM the trend is inverted. Although the feature is not extremely significant it is potentially very interesting.

\begin{figure}[!tp] 
\begin{center}$
\begin{array}{cc}
\includegraphics[width=0.5\columnwidth]{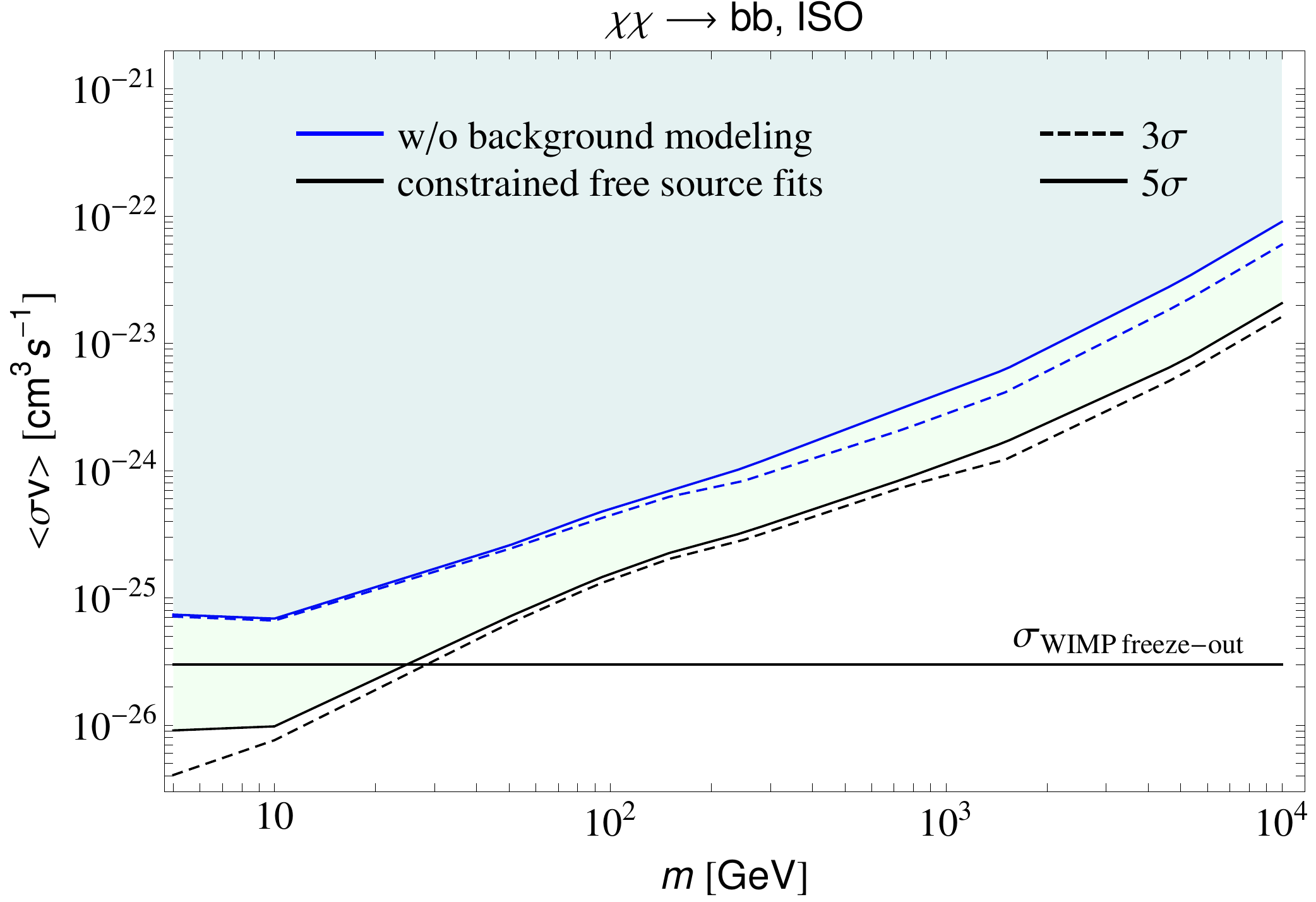}&
\includegraphics[width=0.5\columnwidth]{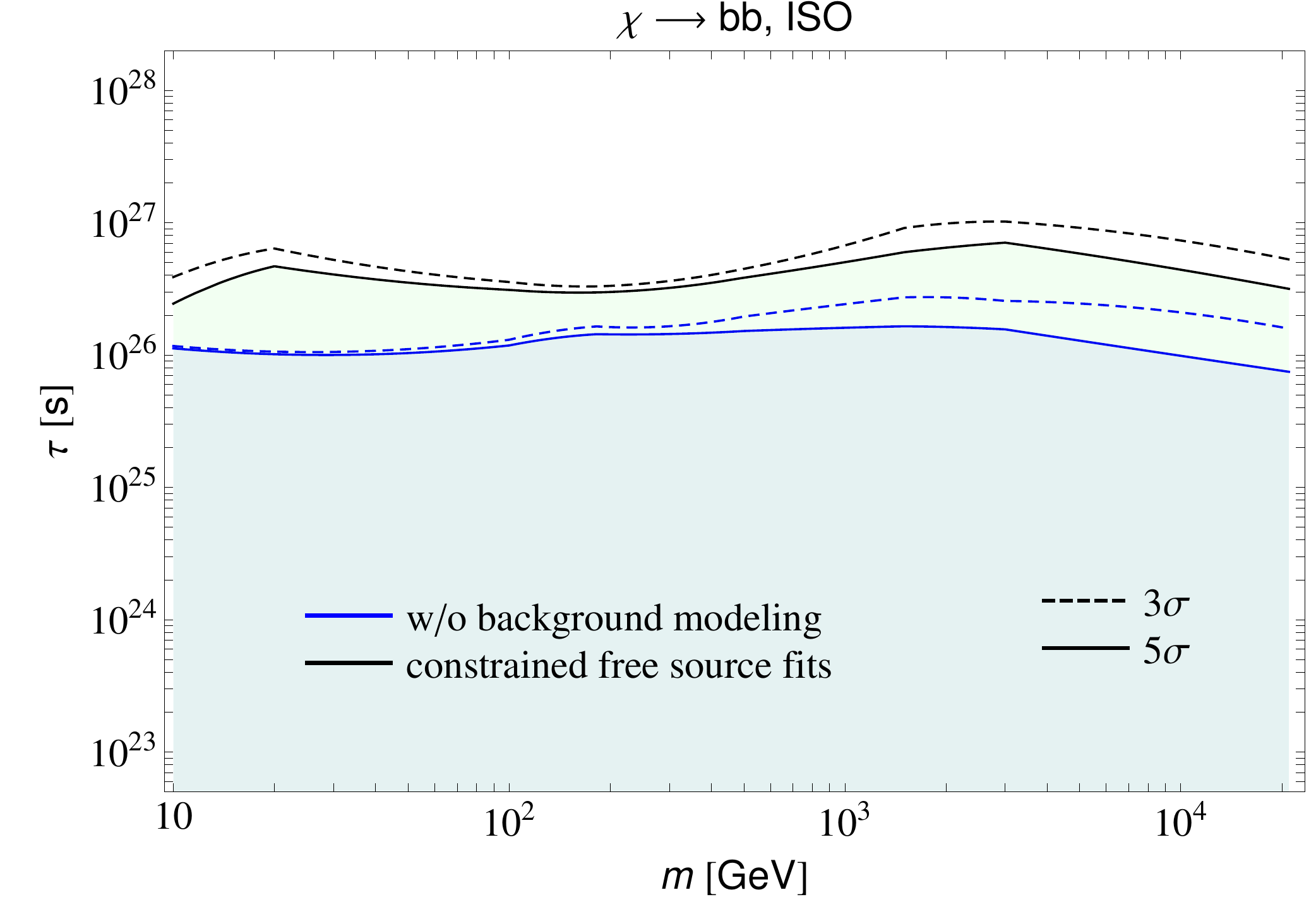}
\end{array}$
\end{center}
%
%
\begin{center}$
\begin{array}{cc}
\includegraphics[width=0.5\columnwidth]{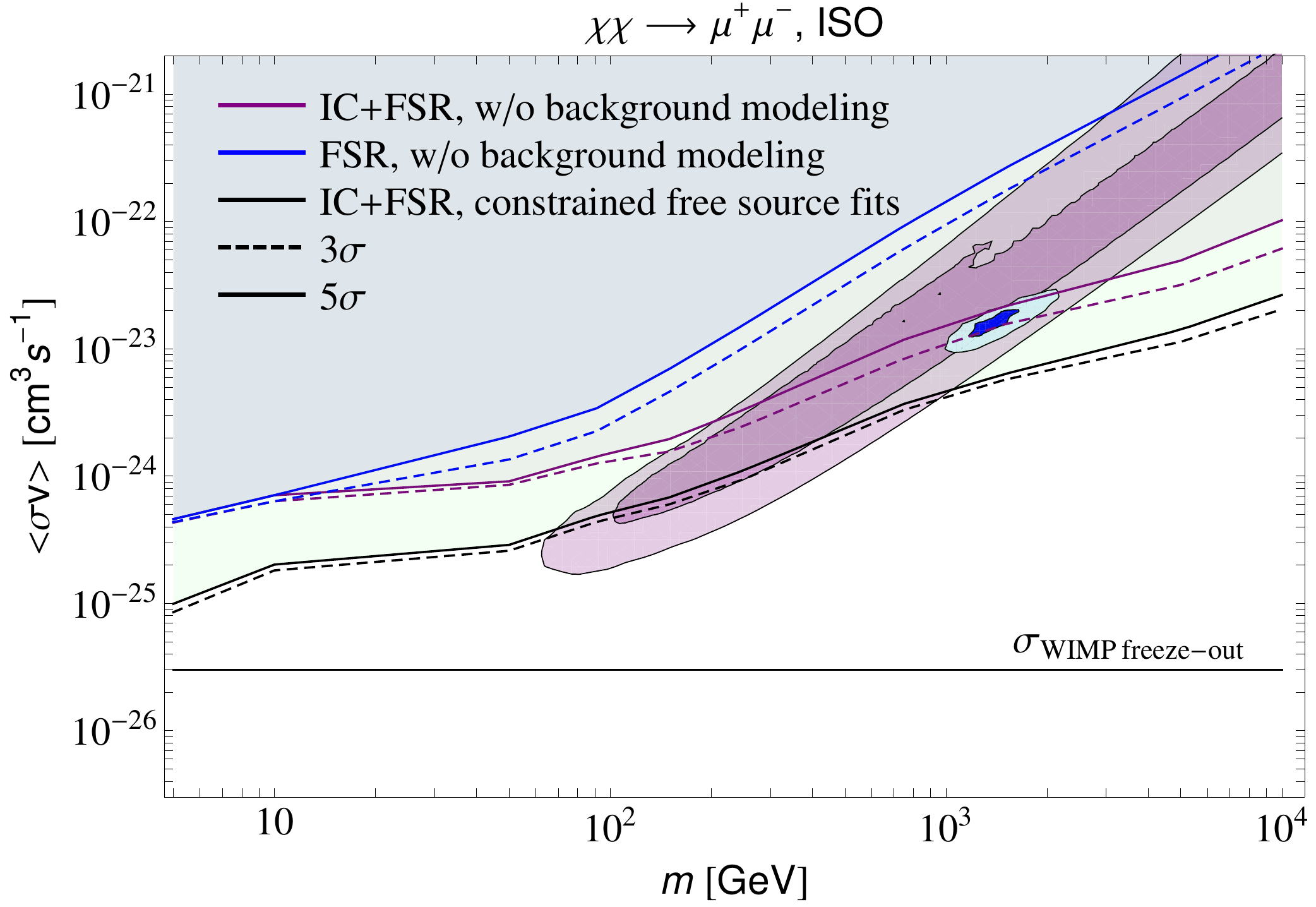}&
\includegraphics[width=0.5\columnwidth]{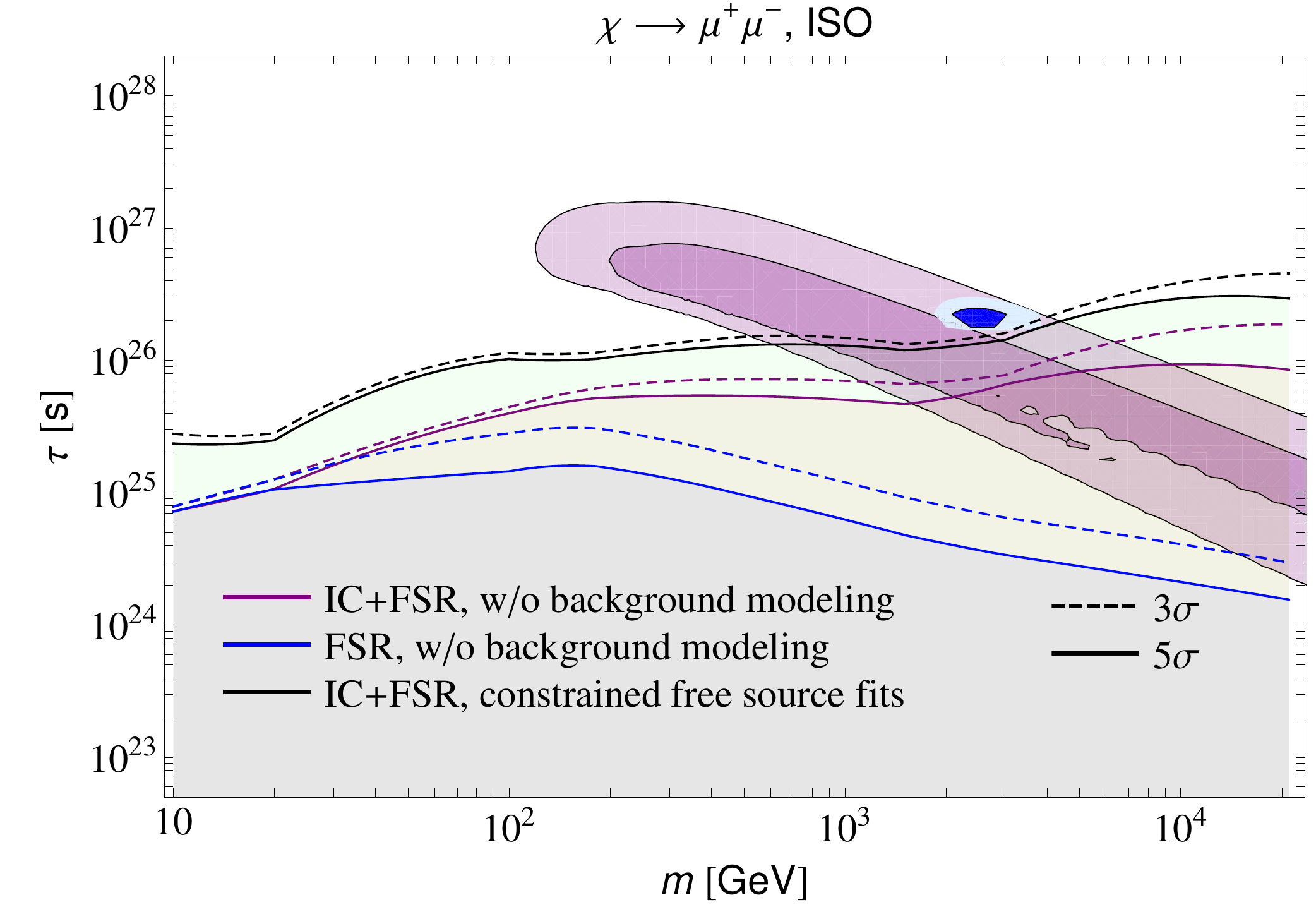}
\end{array}$
\end{center}
\begin{center}$
\begin{array}{cc}
\includegraphics[width=0.5\columnwidth]{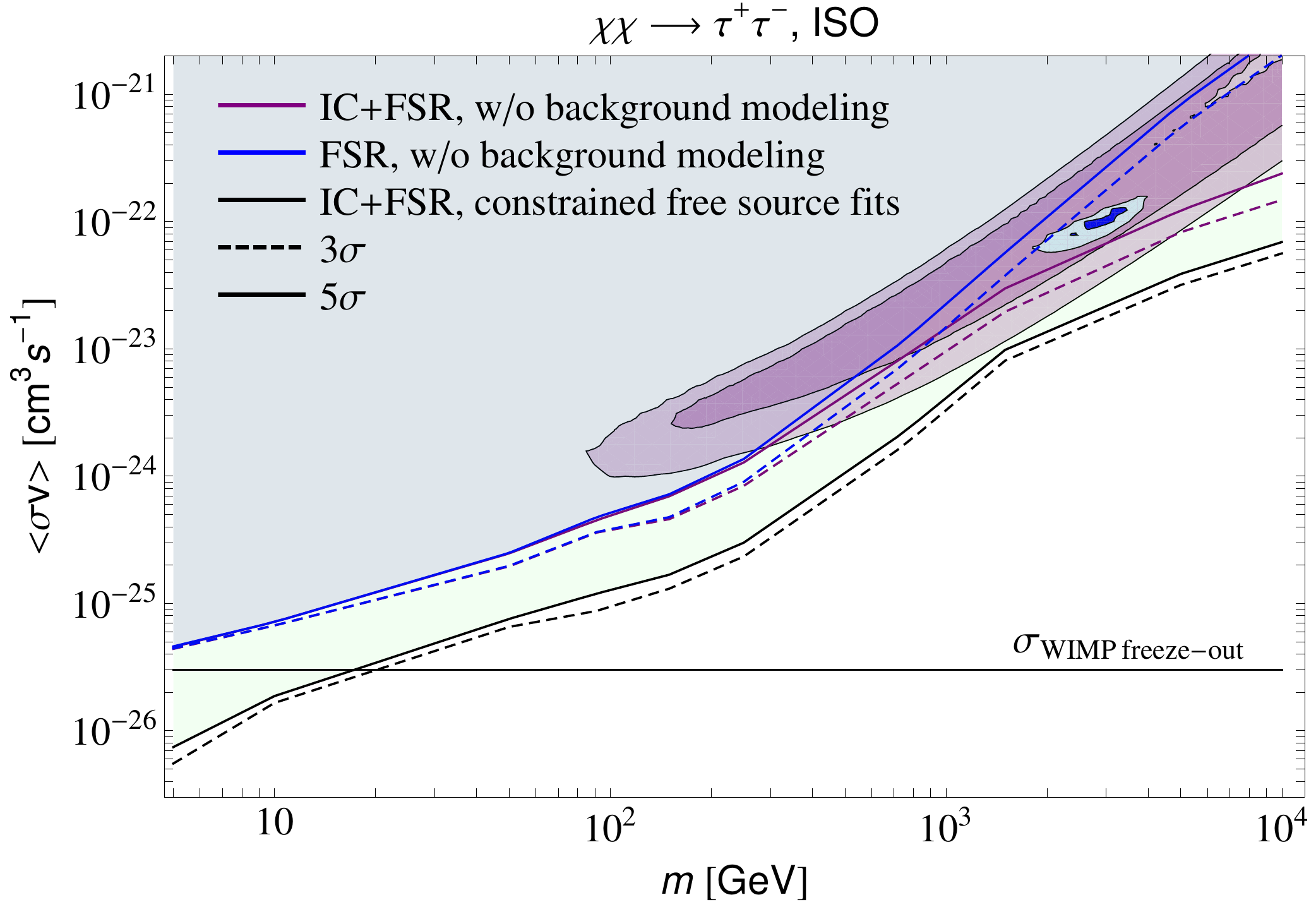}&
\includegraphics[width=0.5\columnwidth]{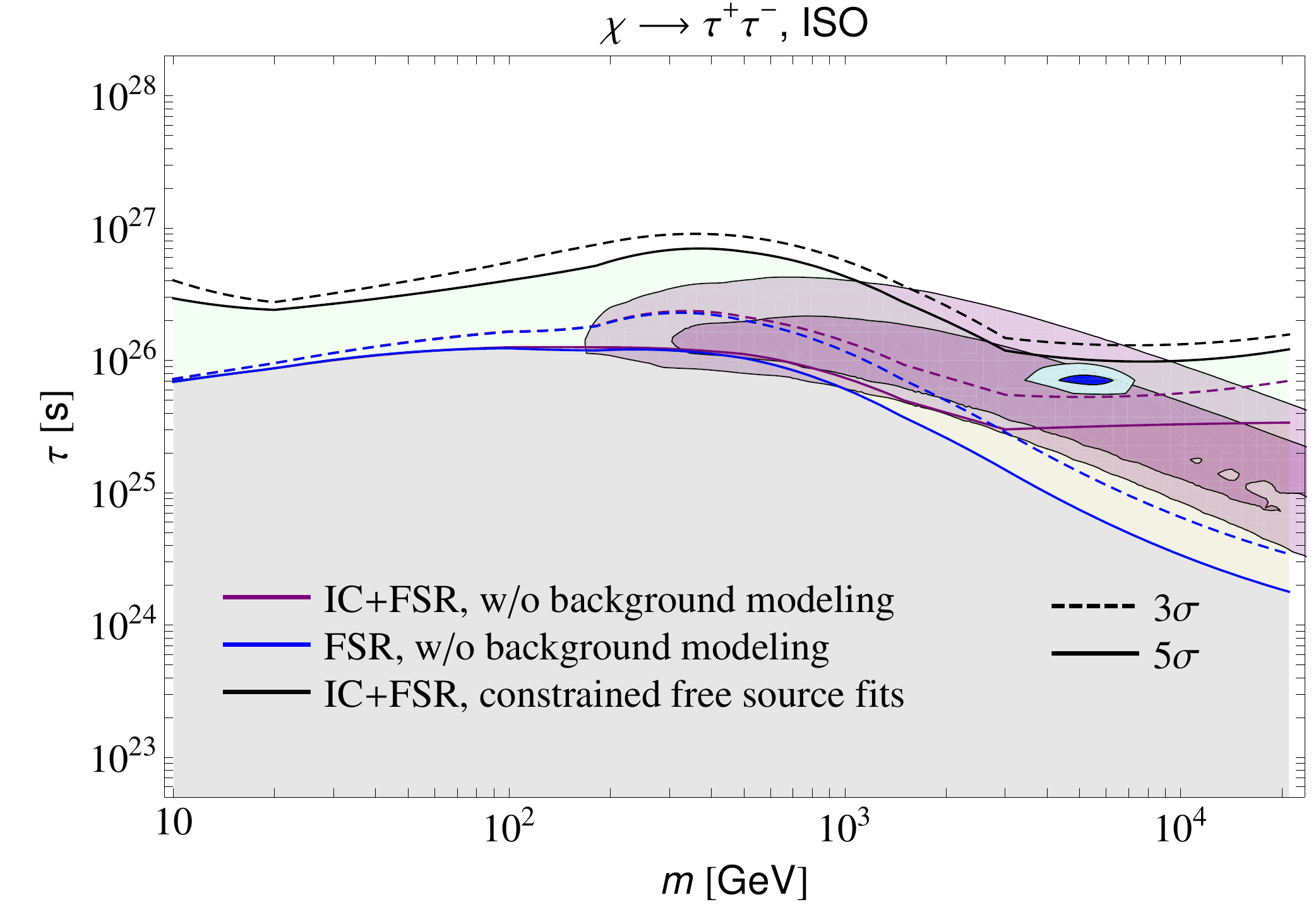}
\end{array}$
\end{center}
\vspace{-0.6cm}
\caption{Upper limits on the velocity averaged DM annihilation cross-section (left) and decay time (right) including a model of the astrophysical background compared with the limits obtained with no modeling of the background. Limits are shown for $b{\bar b}$ (upper), $\mu ^+ \mu^-$ (middle) and $\tau ^+ \tau^-$ (lower panel) channels, for a DM distribution given by the isothermal distribution. The horizontal line marks the thermal decoupling cross section expected for a generic WIMP candidate. The regions of parameter space which provide a good fit to PAMELA (purple) and {\it Fermi} LAT (blue) CR electron and positron data are also shown.  \label{fig:fixedsourcelimits}}
\end{figure}

Upper limits on the velocity averaged annihilation cross section into various channels are shown in Fig.~\ref{fig:fixedsourcelimits}, for isothermal
profile of the DM halo\footnote{Limits obtained using the NFW profile are only slightly better.}, together with regions of parameter space which provide a good fit to PAMELA (purple) and {\it Fermi} LAT (blue) CR electron and positron data \cite{Cirelli:2009dv}. Despite the various conservative choices described above, the resulting DM limits are quite stringent. They are comparable with the limits from LAT searches for a signal from DM annihilation/decay in dwarf galaxies \cite{collaboration:2011wa}. 
In particular, as shown in Figure~\ref{fig:fixedsourcelimits} for masses around 20 GeV the thermal relic value of the annihilation cross section is reached,  both for the $b\bar{b}$  and $\tau^+\tau^-$ channels.  

Overall, rather than being due to residual astrophysical model  uncertainties, the remaining major uncertainties in the DM constraints from the Halo region come from the modeling of the DM signal itself.
The main uncertainty is in the normalization of the DM profile, which is fixed through the local value of the DM density. We use the recent determination $\rho_0=0.43$ GeV cm$^{-3}$ from \cite{Salucci:2010qr}, which has, however, a large uncertainty, with values in the range 0.2-0.7 GeV cm$^{-3}$ still viable. A large uncertainty in $\rho_0$ is particular important for annihilation constraints since they scale like $\rho_0^2$, while for constraints on decaying DM the scaling is only linear. A less important role is played by the uncertainties in the DM profile, since in our region of interest different profiles predict similar DM densities. %
A better determination of the local DM density, as well as of the parameters determining the global structure of the DM Halo, is therefore of the utmost importance for reducing the uncertainties related to DM constraints from DM halo, but it is beyond the scope of this paper and is the subject of dedicated studies.

\newpage
\onecolumn
\begin{table}[t]
\begin{center}
\begin{tabular}{ | c|c|c| }
\hline
\textbf{ {Non linear Parameters}} & \textbf{  Symbol}   & \textbf{  Grid values}  \\
\hline \hline
\small{index of the injection CRE spectrum} & {\bf $\gamma_{e,2}$}   &     \small{1.800, 1.925, 2.050, 2.175, 2.300, 2.425, 2.550, 2.675}  \\
 half height of the diffusive halo\footnote{The parameters $D_0$, $\delta$, $v_A$, $\gamma_{p,1}$, $\gamma_{p,1}$, $\rho_{br,p}$ are varied together with $z_h$ as indicated in Table \ref{pdiffusion}. } &  {\bf $z_h$}   &  2, 4, 6, 8, 10, 15 kpc\\
dust to HI ratio & {\,\, d2HI}  \,\,  &  \small{(0.0120, 0.0130, 0.0140, 0.0150, 0.0160, 0.0170)  $\times 10^{-20}$ mag cm$^2$} \\
\hline
\hline
\textbf{ {Linear Parameters}} & \textbf{  Symbol}   & \textbf{  Range of variation}  \\
\hline \hline
 eCRSD and pCRSD coefficients &  {$c^e_i$,$c^p_i$}  &  0,+$\infty$ \\
 local  H$_2$to CO factor  & {$X_{CO}^{loc}$}  &  0-30 $\times 10^{20}$ cm$^{-2}$  (K km s$^{-1}$)$^{-1}$\\
 IGB normalization in various energy bins & $\alpha_{IGB,m}$  & free \\
 DM normalization &  {$\alpha_{\chi}$}  &  free \\
\hline
\end{tabular}
\caption{Summary table of the parameters varied in the fit. The top part of the table show the non linear parameters and the grid values at which the
likelihood is computed. The bottom part show the linear parameters and the range of variation allowed in the fit. The coefficients of the CRSDs
are forced to be positive, apart  $c^{e,p}_1$ and $c^{e,p}_2$  which are set to zero. The local $X_{CO}$ ratio is restricted to vary in the range 0-30 $\times 10^{20}$ cm$^{-2}$  (K km s$^{-1}$)$^{-1}$, while $\alpha_{IGB,m}$ and $\alpha_{\chi}$ are left free to assume both  positive and negative values. See the text for more details.}
\label{tab:summarypar}
\end{center}
\end{table}

%

\newpage
\onecolumn


\begin{thebibliography}{00}


\bibitem{us}
 M.~Ackermann {\it et al.}  [The Fermi-LAT collaboration],
Astrophys.\ J. in press, 
 	arXiv:1205.6474v1[astro-ph.CO].

\bibitem{paper2}
 M.~Ackermann {\it et al.}  [The Fermi-LAT collaboration],
Astrophys.\ J., {\bf 750}, 3 (2012),
  arXiv:1202.4039 [astro-ph.HE].
  
  
\bibitem{galprop}
  A.~W.~Strong, I.~V.~Moskalenko and O.~Reimer,
  Astrophys.\ J.\  {\bf 537}, 763 (2000),
  url: \emph{http://galprop.stanford.edu/webrun.php}.
  
\bibitem{Casandjian:2009wq}
  J.~-M.~Casandjian, I.~Grenier [The Fermi-LAT collaboration],
  [arXiv:0912.3478 [astro-ph.HE]].


\bibitem{Su:2010qj}
  M.~Su, T.~R.~Slatyer, D.~P.~Finkbeiner,
  Astrophys.\ J.\  {\bf 724}, 1044-1082 (2010). 
  
\bibitem{Navarro:1995iw}
  J.~F.~Navarro, C.~S.~Frenk and S.~D.~M.~White,
  Astrophys.\ J.\  {\bf 462}, 563 (1996).
  
\bibitem{Begeman:1991iy} 
  K.~G.~Begeman, A.~H.~Broeils and R.~H.~Sanders,
  Mon.\ Not.\ Roy.\ Astron.\ Soc.\  {\bf 249}, 523 (1991).
  
\bibitem{Salucci:2010qr}
  P.~Salucci, F.~Nesti, G.~Gentile, C.~F.~Martins,
  Astron.\ Astrophys.\  {\bf 523}, A83 (2010). 


\bibitem{Grasso:2009ma}
  D.~Grasso {\it et al.}  [FERMI-LAT Collaboration],
  Astropart.\ Phys.\  {\bf 32}, 140 (2009).
   P.~Meade, M.~Papucci, A.~Strumia and T.~Volansky,
  Nucl.\ Phys.\  B {\bf 831}, 178 (2010).
  L.~Bergstrom, J.~Edsjo and G.~Zaharijas,
  Phys.\ Rev.\ Lett.\  {\bf 103}, 031103 (2009).


  
  \bibitem{Adriani:2008zr}
  O.~Adriani {\it et al.}  [PAMELA Collaboration],
  Nature {\bf 458}, 607 (2009).

\bibitem{Abdo:2009zk}
  A.~A.~Abdo {\it et al.}  [The Fermi LAT Collaboration],
  Phys.\ Rev.\ Lett.\  {\bf 102}, 181101 (2009).
  
\bibitem{Cirelli:2010xx}
  M.~Cirelli {\it et al.},
  JCAP {\bf 1103} (2011) 051.

\bibitem{1FGL}
  A.~A.~Abdo {\it et al.}  [The Fermi-LAT collaboration],
  Astrophys.\ J.\ Suppl.\  {\bf 188 } (2010)  405-436.


\bibitem{Cirelli:2009dv}
  M.~Cirelli, P.~Panci, P.~D.~Serpico,
  Nucl.\ Phys.\  {\bf B840 } (2010)  284-303.
  
\bibitem{collaboration:2011wa} 
  M.~Ackermann {\it et al.}  [Fermi-LAT Collaboration],
  Phys.\ Rev.\ Lett.\  {\bf 107}, 241302 (2011).




 \end{thebibliography}
 \end{document}